\documentclass[twocolumn,showpacs,preprintnumbers,amsmath,amssymb,prl]{revtex4}

\usepackage{graphicx}
\usepackage{dcolumn}
\usepackage{bm}
\usepackage{tabularx}
\usepackage{amsmath}

\begin{document}

\title{Nodal superconductivity in Ba(Fe$_{1-x}$Ru$_x$)$_2$As$_2$ induced by isovalent Ru substitution}

\author{X. Qiu,$^1$ S. Y. Zhou,$^1$ H. Zhang,$^1$ B. Y. Pan,$^1$ X. C. Hong,$^1$ Y. F. Dai,$^1$ Man Jin Eom,$^2$ Jun Sung Kim,$^2$ S. Y. Li$^{1,*}$}

\affiliation{$^1$Department of Physics, State Key Laboratory of Surface Physics, and Laboratory of Advanced Materials, Fudan University, Shanghai 200433, China\\
$^2$Department of Physics, Pohang University of Science and
Technology, Pohang 790-784, Korea}

\date{\today}

\begin{abstract}
We present the ultra-low-temperature heat transport study of an
iron-based superconductor Ba(Fe$_{0.64}$Ru$_{0.36}$)$_2$As$_2$
($T_c$ = 20.2 K), in which the superconductivity is induced by
isovalent Ru substitution. In zero field we find a large residual
linear term $\kappa_0/T$, more than 40\% of the normal-state value.
At low field, the $\kappa_0/T$ shows an $H^{1/2}$ dependence. These
provide strong evidences for nodes in the superconducting gap of
Ba(Fe$_{0.64}$Ru$_{0.36}$)$_2$As$_2$, which mimics that in another
isovalently substituted superconductor
BaFe$_2$(As$_{1-x}$P$_x$)$_2$. Our results show that the isovalent
Ru substitution can also induce nodal superconductivity in
BaFe$_2$As$_2$, as P does, and they may have the same origin. We
further compare them with other two nodal superconductors LaFePO and
LiFeP.
\end{abstract}


\maketitle

Since the discovery of high-$T_c$ superconductivity in iron-based
compounds \cite{Kamihara,Paglione}, the electronic pairing mechanism
has been a central issue \cite{FaWang1}. One key to understand it is
to clarify the symmetry and structure of the superconducting gap
\cite{Hirschfeld}. However, even for the most studied
(Ba,Sr,Ca,Eu)Fe$_2$As$_2$ (122) system, the situation is still
fairly complex \cite{Hirschfeld}.

Near optimal doping, for both hole- and electron-doped 122
compounds, the angle-resolved photon emission spectroscopy (ARPES)
experiments clearly demonstrated multiple nodeless superconducting
gaps \cite{HDing,KTerashima}, which was further supported by bulk
measurements such as thermal conductivity
\cite{XGLuo,LDing,Tanatar}. On the overdoped side, nodal
superconductivity was found in the extremely hole-doped
KFe$_2$As$_2$ \cite{JKDong1,KHashimoto1}, while strongly anisotropic
gap \cite{Tanatar}, or isotropic gaps with significantly different
magnitude \cite{JKDong2,YBang} were suggested in the heavily
electron-doped Ba(Fe$_{1-x}$Co$_x$)$_2$As$_2$. On the underdoped
side, recent heat transport measurements claimed possible nodes in
the superconducting gap of hole-doped Ba$_{1-x}$K$_x$Fe$_2$As$_2$
with $x <$ 0.16 \cite{JReid1}, in contrast to the nodeless gaps
found in electron-doped Ba(Fe$_{1-x}$Co$_x$)$_2$As$_2$
\cite{Tanatar}.

Intriguingly, nodal superconductivity was also found in
BaFe$_2$(As$_{1-x}$P$_x$)$_2$ ($T_c =$ 30 K)
\cite{YNakai,KHashimoto2}, in which the superconductivity is induced
by the isovalent P substitution for As. The laser-ARPES experiments
on BaFe$_2$(As$_{0.65}$P$_{0.35}$)$_2$ showed isotropic gaps in the
three hole pockets around the Brillouin zone (BZ) center, therefore
the gap nodes must locate on the electron pockets around the BZ
corners \cite{TShimojima}. Moreover, previously LaFePO ($T_c \sim$ 6
K) displays clear nodal behavior \cite{Fletcher,Hicks,MYamashida},
and recently there is penetration depth evidence for nodes in the
superconducting gap of LiFeP ($T_c \sim$ 4.5 K) \cite{KHashimoto3}.
The nodal superconductivity in these P-substituted compounds are
very striking, which raises the puzzling question why the P
substitution is so special in iron-based superconductors. The
theoretical explanations of this puzzle are far from consensus
\cite{KKuroki,FaWang2,RThomale,KSuzuki}.

In the Fe$_2$As$_2$ slabs of iron arsenides shown in Fig. 1, instead
of substituting As with P, there is an alternative way for isovalent
substitution, to substitute Fe with Ru. Indeed, superconductivity
with $T_c$ up to 20 K was found in Ba(Fe$_{1-x}$Ru$_x$)$_2$As$_2$
\cite{SSharma} and Sr(Fe$_{1-x}$Ru$_x$)$_2$As$_2$ \cite{WSchnelle}.
The phase diagram of Ba(Fe$_{1-x}$Ru$_x$)$_2$As$_2$ is very similar
to that of BaFe$_2$(As$_{1-x}$P$_x$)$_2$
\cite{FRullier-Albenque,AThaler}. The ARPES measurements on
Ba(Fe$_{0.65}$Ru$_{0.35}$)$_2$As$_2$ showed that Ru induces neither
hole nor electron doping, but the hole and electron pockets are
about twice larger than in BaFe$_2$As$_2$ \cite{VBrouet}. To
investigate the superconducting gap structure in these
Ru-substituted superconductors may help to solve above puzzle of P
substitution.

In this Letter, we report the demonstration of nodal
superconductivity in optimally substituted
Ba(Fe$_{0.64}$Ru$_{0.36}$)$_2$As$_2$ by thermal conductivity
measurements down to 50 mK. Our finding shows that the nodal
superconducting states in P-substituted iron arsenides are not that
special, and suggests a common origin of the nodal superconductivity
induced by isovalent substitutions, at least in
Ba(Fe$_{1-x}$Ru$_x$)$_2$As$_2$ and BaFe$_2$(As$_{1-x}$P$_x$)$_2$.

Single crystals of optimally substituted
Ba(Fe$_{0.64}$Ru$_{0.36}$)$_2$As$_2$ were grown using a self-flux
method \cite{ManJinEom}. Plate-shaped crystals with shiny surfaces
were extracted mechanically. The Ru substituting level was
determined by energy dispersive X-ray spectroscopy. The dc magnetic
susceptibility was measured at $H$ = 10 Oe, with zero-field cooled,
using a SQUID (MPMS, Quantum Design). The sample was cleaved to a
rectangular shape of dimensions 1.50 $\times$ 0.68 mm$^2$ in the
$ab$-plane, with 70 $\mu$m thickness along the $c$-axis. Contacts
were made directly on the sample surfaces with silver paint, which
were used for both resistivity and thermal conductivity
measurements. The contacts are metallic with typical resistance 200
m$\Omega$ at 1.5 K. In-plane thermal conductivity was measured in a
dilution refrigerator, using a standard four-wire steady-state
method with two RuO$_2$ chip thermometers, calibrated {\it in situ}
against a reference RuO$_2$ thermometer. Magnetic fields were
applied along the $c$-axis. To ensure a homogeneous field
distribution in the sample, all fields were applied at temperature
above $T_c$.

\begin{figure}
\includegraphics[clip,width=4cm]{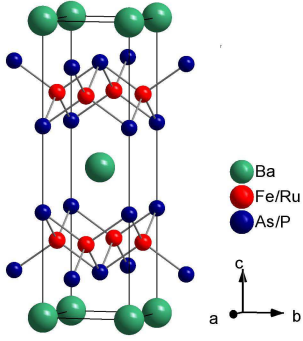}
\caption{(Color online). Crystal structure of BaFe$_2$As$_2$. There
are two ways for isovalent substitution in the Fe$_2$As$_2$ slabs,
substituting As with P, or Fe with Ru. Both substitutions can induce
superconductivity, and result in similar phase diagrams.}
\end{figure}

\begin{figure}
\includegraphics[clip,width=5.2cm]{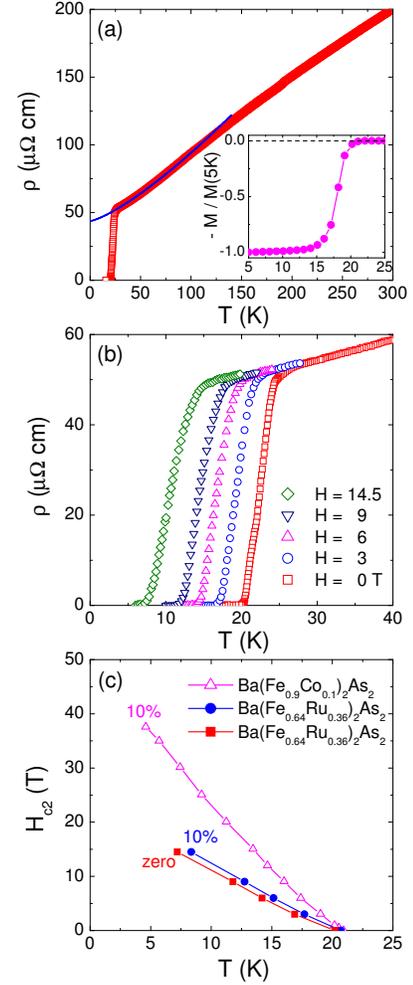}
\caption{(Color online). (a) In-plane resistivity of
Ba(Fe$_{0.64}$Ru$_{0.36}$)$_2$As$_2$ single crystal. The solid line
is a fit of the data between 30 and 90 K to $\rho(T) = \rho_0
+AT^n$. Inset: normalized magnetization. (b) Low-temperature
resistivity in $H$ = 0, 3, 6, 9, and 14.5 T with $H || c$. (c)
Temperature dependence of the upper critical field $H_{c2}$. The
squares and circles represent $H_{c2}$ of
Ba(Fe$_{0.64}$Ru$_{0.36}$)$_2$As$_2$ defined by $\rho = 0$ and $\rho
= 0.1\rho_N$, respectively. The triangles represent $H_{c2}$ of
Ba(Fe$_{0.9}$Co$_{0.1}$)$_2$As$_2$ with $T_c \approx 22$ K
\cite{AYamamoto}.}
\end{figure}

Fig. 2a shows the in-plane resistivity $\rho(T)$ of our
Ba(Fe$_{0.64}$Ru$_{0.36}$)$_2$As$_2$ single crystal. The magnitude
and shape of $\rho(T)$ are consistent with previous report
\cite{FRullier-Albenque}. The normalized magnetization was plotted
in the inset, which displays a nice superconducting transition at
about 20 K. According to the phase diagram of
Ba(Fe$_{1-x}$Ru$_x$)$_2$As$_2$ \cite{FRullier-Albenque,AThaler}, no
static magnetic order exists in our optimally substituted sample.
The resistivity data between 30 and 90 K are fitted to $\rho(T) =
\rho_0 +AT^n$, which gives a residual resistivity $\rho_0$ = 43.7
$\pm$ 0.1 $\mu\Omega$cm and $n$ = 1.31 $\pm$ 0.1. Such a
non-Fermi-liquid temperature dependence of $\rho(T)$ is similar to
that observed in BaFe$_2$(As$_{1-x}$P$_x$)$_2$ near optimal
substitution, which may reflet the presence of antiferromagnetic
spin fluctuations near a quantum critical point \cite{SKasahara}.

The low-temperature part of $\rho(T)$ is plotted in Fig. 2b. The
zero-resistance point of the resistive transition is at $T_c$ = 20.2
K, which is in good agreement with the diamagnetic superconducting
transition shown in the inset of Fig. 2a. To estimate the upper
critical field $H_{c2}$, the resistivity in $H$ = 3, 6, 9, and 14.5
T with $H || c$ were also measured, as seen in Fig. 2b. Fig. 2c
shows the temperature dependence of $H_{c2}$, defined by $\rho = 0$
and $\rho = 0.1\rho_N$, respectively. For comparison, the
$H_{c2}(T)$ of the electron-doped Ba(Fe$_{0.9}$Co$_{0.1}$)$_2$As$_2$
single crystal with $T_c \approx$ 22 K is reproduced from ref.
\cite{AYamamoto}. One can see that although the $T_c$ of our
Ba(Fe$_{0.64}$Ru$_{0.36}$)$_2$As$_2$ is only slightly lower than
Ba(Fe$_{0.9}$Co$_{0.1}$)$_2$As$_2$, its $H_{c2}(T)$ is significantly
lower. For Ba(Fe$_{0.9}$Co$_{0.1}$)$_2$As$_2$, extrapolation of the
$H_{c2}(T)$ data suggests $H_{c2}(0)$ between 40 and 50 T, much
larger than that obtained from Werthamer-Helfand-Hohenberg formula
$H_{c2}^{WHH}(0) = -0.69T_cdH_{c2}/dT|_{T=T_c}$ \cite{AYamamoto}.
For our Ba(Fe$_{0.64}$Ru$_{0.36}$)$_2$As$_2$, $H_{c2}^{WHH}(0)
\approx$ 15.3 T is obtained, which is also apparently lower than the
actual $H_{c2}(0)$. We can only roughly estimate the bulk $H_{c2}(0)
\approx$ 23 T, defined by $\rho = 0$, by linearly extrapolating the
data between 6 and 14.5 T in Fig. 2c. Note that a slightly different
$H_{c2}(0)$ does not affect our discussion below.

Fig. 3 shows the temperature dependence of the in-plane thermal
conductivity for Ba(Fe$_{0.64}$Ru$_{0.36}$)$_2$As$_2$ in $H$ = 0, 1,
2, 4, 6, 9, and 12 T magnetic fields, plotted as $\kappa/T$ vs $T$.
All the curves are roughly linear, as previously observed in
BaFe$_{1.9}$Ni$_{0.1}$As$_2$ \cite{LDing}, KFe$_2$As$_2$
\cite{JKDong1}, and overdoped Ba(Fe$_{1-x}$Co$_x$)$_2$As$_2$ single
crystals \cite{Tanatar,JKDong2}. Therefore we fit all the curves to
$\kappa/T$ = $a + bT^{\alpha-1}$ with $\alpha$ fixed to 2. The two
terms $aT$ and $bT^{\alpha}$ represent contributions from electrons
and phonons, respectively. Here we only focus on the electronic
term.

In zero field, the fitting gives a residual linear term $\kappa_0/T$
= 0.266 $\pm$ 0.002 mW K$^{-2}$ cm$^{-1}$. This value is more than
40\% of the normal-state Wiedemann-Franz law expectation
$\kappa_{N0}/T$ = $L_0$/$\rho_0$ = 0.56 mW K$^{-2}$ cm$^{-1}$, with
$L_0$ the Lorenz number 2.45 $\times$ 10$^{-8}$ W$\Omega$K$^{-2}$
and $\rho_0$ = 43.7 $\mu\Omega$cm. For another isovalently
substituted BaFe$_2$(As$_{0.67}$P$_{0.33}$)$_2$ single crystal,
similar value of $\kappa_0/T \approx$ 0.25 mW K$^{-2}$ cm$^{-1}$ was
obtained, which is about 30\% of its normal-state $\kappa_{N0}/T$
\cite{KHashimoto2}. The significant $\kappa_0/T$ of
Ba(Fe$_{0.64}$Ru$_{0.36}$)$_2$As$_2$ in zero field is a strong
evidence for nodes in the superconducting gap \cite{Shakeripour}.

\begin{figure}
\includegraphics[clip,width=6.5cm]{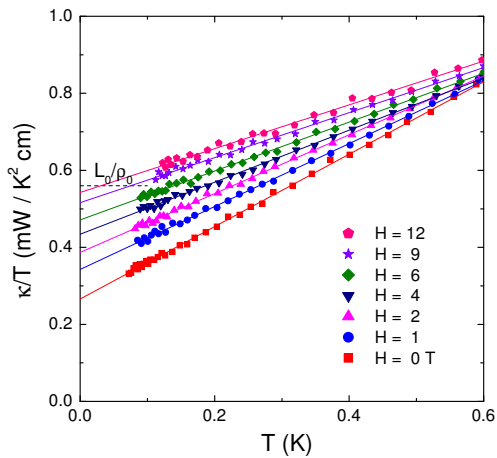}
\caption{(Color online). Low-temperature in-plane thermal
conductivity of Ba(Fe$_{0.64}$Ru$_{0.36}$)$_2$As$_2$ in magnetic
fields applied along the $c$-axis ($H$ = 0, 1, 2, 4, 6, 9 and 12 T).
The solid lines are $\kappa/T = a + bT$ fit to all the curves,
respectively. The dash line is the normal-state Wiedemann-Franz law
expectation $L_0$/$\rho_0$, with $L_0$ the Lorenz number 2.45
$\times$ 10$^{-8}$ W$\Omega$K$^{-2}$ and $\rho_0$ = 43.7
$\mu\Omega$cm.}
\end{figure}

\begin{figure}
\includegraphics[clip,width=6.45cm]{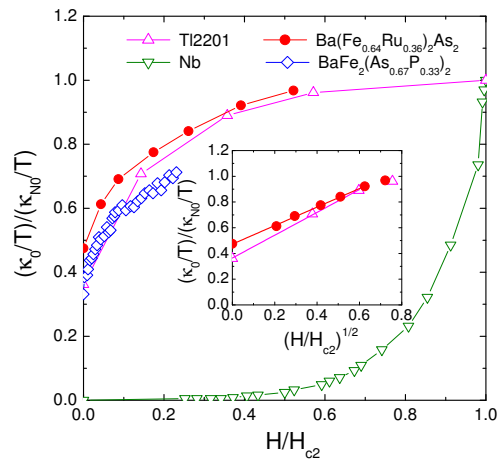}
\caption{(Color online). Normalized residual linear term
$\kappa_0/T$ of Ba(Fe$_{0.64}$Ru$_{0.36}$)$_2$As$_2$ as a function
of $H/H_{c2}$. Similar data of the clean $s$-wave superconductor Nb
\cite{Lowell}, an overdoped $d$-wave cuprate superconductor Tl-2201
\cite{Proust}, and BaFe$_2$(As$_{0.67}$P$_{0.33}$)$_2$
\cite{KHashimoto2} are also shown for comparison. The behavior of
$\kappa_0(H)/T$ in Ba(Fe$_{0.64}$Ru$_{0.36}$)$_2$As$_2$ clearly
mimics that in Tl-2201 and BaFe$_2$(As$_{0.67}$P$_{0.33}$)$_2$.
Inset: the same data of Ba(Fe$_{0.64}$Ru$_{0.36}$)$_2$As$_2$ and
Tl-2201 plotted against $(H/H_{c2})^{1/2}$. The lines represent the
$H^{1/2}$ dependence.}
\end{figure}

The field dependence of $\kappa_0/T$ may provide further support for
the nodes \cite{Shakeripour}. In Fig. 4, the normalized
$(\kappa_0/T)/(\kappa_{N0}/T)$ of
Ba(Fe$_{0.64}$Ru$_{0.36}$)$_2$As$_2$ is plotted as a function of
$H/H_{c2}$, with the normal-state $\kappa_{N0}/T$ = 0.56 mW K$^{-2}$
cm$^{-1}$ and bulk $H_{c2}$ = 23 T. Similar data of the clean
$s$-wave superconductor Nb \cite{Lowell}, an overdoped $d$-wave
cuprate superconductor Tl-2201 \cite{Proust}, and
BaFe$_2$(As$_{0.67}$P$_{0.33}$)$_2$ \cite{KHashimoto2} are also
plotted for comparison. For a nodal superconductor in magnetic
field, delocalized states exist out the vortex cores and dominate
the heat transport in the vortex state, in contrast to the $s$-wave
superconductor. At low field, the Doppler shift due to superfluid
flow around the vortices will yield an $H^{1/2}$ growth in
quasiparticle density of states (the Volovik effect \cite{Volovik}),
thus the $H^{1/2}$ field dependence of $\kappa_0/T$. From Fig. 4,
the behavior of $\kappa_0(H)/T$ in
Ba(Fe$_{0.64}$Ru$_{0.36}$)$_2$As$_2$ clearly mimics that in Tl-2201
and BaFe$_2$(As$_{0.67}$P$_{0.33}$)$_2$. In the inset of Fig. 4, the
$\kappa_0(H)/T$ of Ba(Fe$_{0.64}$Ru$_{0.36}$)$_2$As$_2$ obeys the
$H^{1/2}$ dependence at low field, which supports the existence of
nodes in the superconducting gap.

To our knowledge, previously there are five iron-based
superconductors displaying nodal superconductivity, KFe$_2$As$_2$
\cite{JKDong1,KHashimoto1}, underdoped Ba$_{1-x}$K$_x$Fe$_2$As$_2$
($x <$ 0.16) \cite{JReid1}, BaFe$_2$(As$_{1-x}$P$_x$)$_2$
\cite{YNakai,KHashimoto2}, LaFePO \cite{Fletcher,Hicks,MYamashida},
and LiFeP \cite{KHashimoto3}. Here we only consider the ``in-plane
nodes", not counting the ``$c$-axis nodes" in underdoped and
overdoped Ba(Fe$_{1-x}$Co$_x$)$_2$As$_2$ as suggested by $c$-axis
heat transport experiments \cite{JReid2}. For the extremely
hole-doped KFe$_2$As$_2$, the nodal superconductivity may result
from the intraband pairing via antiferromagnetic fluctuations, due
to the lack of electron pockets \cite{JKDong1}. For underdoped
Ba$_{1-x}$K$_x$Fe$_2$As$_2$, it is still not clear how the
superconducting gap transforms from nodeless to nodal at $x \approx
0.16$ \cite{JReid1}. The rest three compounds,
BaFe$_2$(As$_{1-x}$P$_x$)$_2$, LaFePO, and LiFeP, have stimulated
various interpretations of the effect of isovalent P substitution on
the superconducting gap structure
\cite{KKuroki,FaWang2,RThomale,KSuzuki}.

Our new finding of nodal superconductivity in
Ba(Fe$_{0.64}$Ru$_{0.36}$)$_2$As$_2$ reveals the similarity between
the isovalently Ru- and P-substituted iron arsenides, therefore the
P substitution is not that special for inducing nodal
superconductivity. In this sense, the mystery of P substitution in
iron-based superconductors has been partially unwrapped. What next
one needs to do is to find out whether there is a common origin for
the nodal superconductivity in these isovalently substituted
compounds.

Due to the smaller size of P ion than As ion, one common structural
feature of the P-substituted compounds is the decrease of pnictogen
height and increase of As-Fe-As angle
\cite{MTegel,SKasahara,SJiang}. The substitution of larger Ru ion
for Fe ion in Ba(Fe$_{1-x}$Ru$_x$)$_2$As$_2$ results in the increase
of $a$ lattice parameter and decrease of $c$ lattice parameter, thus
the decrease of pnictogen height and increase of As-Fe-As angle too
\cite{FRullier-Albenque}. Therefore, both the P and Ru substitutions
cause the same trend of structure change in iron arsenides.

LaFePO and LiFeP belong to the ``1111" and ``111" systems,
respectively, and the P ions have fully substituted As ions in
LaFeAsO and LiFeAs. This is different from the partial P and Ru
substitution in superconducting BaFe$_2$(As$_{1-x}$P$_x$)$_2$ and
Ba(Fe$_{1-x}$Ru$_x$)$_2$As$_2$. In fact, both the fully substituted
BaFe$_2$P$_2$ and BaRu$_2$As$_2$ are nonsuperconducting
\cite{HShishido,RNath}. Another difference is the much lower $T_c$
of LaFePO and LiFeP, 6 and 4.5 K, respectively. For LaFePO, Kuroki
{\it et al.} have attributed the low-$T_c$ nodal pairing to the lack
of Fermi surface $\gamma$ around ($\pi,\pi$) in the unfolded
Brillouin zone, due to the low pnictogen height \cite{KKuroki}.
Hashimoto {\it et al.} also related the nodal superconductivity in
LiFeP to the pnictogen height \cite{KHashimoto3}.

For Ba(Fe$_{1-x}$Ru$_x$)$_2$As$_2$ and
BaFe$_2$(As$_{1-x}$P$_x$)$_2$, the substitutions start from the same
parent compound BaFe$_2$As$_2$, and result in similar phase diagrams
\cite{FRullier-Albenque,AThaler}. The highest $T_c$ at optimal
substitution, 20 and 30 K, are also close. Furthermore, the Fermi
surface structures are roughly similar, with hole pockets around BZ
center and electron pockets around BZ corners
\cite{TShimojima,VBrouet}. All these similarities suggest that the
origin of the nodal superconductivity in
Ba(Fe$_{1-x}$Ru$_x$)$_2$As$_2$ and BaFe$_2$(As$_{1-x}$P$_x$)$_2$ may
be the same. Suzuki {\it et al.} have proposed three-dimensional
nodal structure in the largely warped hole Fermi surface and no
nodes on the electron Fermi surface \cite{KSuzuki}. However, these
seem inconsistent with the ARPES results, which have constrained the
nodes on the electron pockets \cite{TShimojima}. More careful
considerations of the structural parameters, band structure, and
local interactions are needed to clarify whether there is a common
origin for the nodal superconductivity in all these isovalently
substituted iron arsenides.

In summary, we have measured the thermal conductivity of
Ba(Fe$_{0.64}$Ru$_{0.36}$)$_2$As$_2$ single crystal down to 50 mK. A
large $\kappa_0/T$ at zero field and an $H^{1/2}$ field dependence
of $\kappa_0(H)/T$ at low field give strong evidences for nodal
superconductivity in Ba(Fe$_{0.64}$Ru$_{0.36}$)$_2$As$_2$. Comparing
with previous P-substituted iron arsenides, our new finding suggest
that the nodal superconductivity induced by isovalent substitutions
may have the same origin, at least in Ba(Fe$_{1-x}$Ru$_x$)$_2$As$_2$
and BaFe$_2$(As$_{1-x}$P$_x$)$_2$. Finding out this origin will be
important for getting a complete electronic pairing mechanism of the
iron-based high-$T_c$ superconductors.

This work is supported by the Natural Science Foundation of China,
the Ministry of Science and Technology of China (National Basic
Research Program No: 2009CB929203), Program for New Century
Excellent Talents in University, Program for Professor of Special
Appointment (Eastern Scholar) at Shanghai Institutions of Higher
Learning, and STCSM of China (No: 08dj1400200 and 08PJ1402100). \\

$^*$ E-mail: shiyan$\_$li@fudan.edu.cn

\end{document}